\RequirePackage[table,dvipsnames]{xcolor}
\documentclass{webofc}
\usepackage[varg]{txfonts}   
\usepackage{xspace}
\usepackage{cleveref}
\usepackage[inkscape=newer,inkscapeformat=png]{svg}
\usepackage{subcaption}
\usepackage{siunitx}
\DeclareSIUnit\permille{\text{\textperthousand}}
\usepackage{adjustbox}
\usepackage{booktabs}

\newcommand{\mgfamc}{Madgraph5\_aMC@NLO\xspace}
\newcommand{\mg}{MG5\_aMC\xspace}
\newcommand{\mgfgpu}{madgraph4gpu\xspace}
\newcommand{\cpp}{C++\xspace}
\newcommand{\process}[2]{$ #1 \, {\rightarrow} \, #2 $}
\newcommand{\x}{\footnotesize \ensuremath{\times} }
\newcommand{\eemumu}{\process{e^+\!e^-}{\mu^+\!\mu^-}}
\DeclareSIUnit[]{\pb}{pb}
\begin{document}
\title{\mgfamc on GPUs and vector CPUs}
%
%
\subtitle{Experience with the first alpha release}

\author{
\firstname{Stephan} \lastname{Hageb\"ock}    \inst{1}\fnsep\thanks{\email{stephan.hageboeck@cern.ch}} \and
\firstname{Taylor}  \lastname{Childers}     \inst{2} \and
\firstname{Walter}  \lastname{Hopkins}      \inst{2} \and
\firstname{Olivier} \lastname{Mattelaer}    \inst{3} \and
\firstname{Nathan}  \lastname{Nichols}      \inst{2} \and
\firstname{Stefan}  \lastname{Roiser}       \inst{1} \and
\firstname{J{\o}rgen}  \lastname{Teig}         \inst{1}  \and
\firstname{Andrea}  \lastname{Valassi}      \inst{1} \and
\firstname{Carl}    \lastname{Vuosalo}      \inst{4}  \and
\firstname{Zenny}   \lastname{Wettersten}   \inst{1}
}

\institute{CERN, Geneva, Switzerland
\and
    Argonne National Laboratory, Illinois, USA
\and
    Universit\'e Catholique de Louvain, Louvain, Belgium
\and
    University of Wisconsin-Madison, Madison, USA
}

\abstract{%
  \mgfamc is one of the most-frequently used Monte-Carlo event generators at the LHC, and an important consumer of compute resources. The software has been reengineered to maintain the overall look-and-feel of the user interface while speeding up event generation on CPUs and GPUs. The most computationally intensive part, the calculation of ``matrix elements'', is offloaded to new implementations optimised for GPUs and for CPU vector instructions, using event-level data parallelism. We present the work to support accelerated leading-order QCD processes, and discuss how this work is going to be released to \mgfamc's users.
}
\maketitle
\section{Introduction}
\label{sec:intro}

As a Monte-Carlo event generator, the \mgfamc framework (\mg)~\cite{madgraph2014} stands at the beginning of the simulation chain for experiments at the Large Hadron Collider (LHC)~\cite{LHC} or other colliders.
Event generation, and the subsequent simulation of the interaction of generated particles with the detectors account for almost half of the computing resources spent by the LHC experiments~\cite{CERN-LHCC-2022-005,CMSComputing2022}.
With longer operation time of the LHC, the collision data recorded by the experiments is increasing, and a corresponding increase in simulated events is desirable to conduct high-precision analyses.
At the same time, the computing resources of the Worldwide LHC Computing Grid (WLCG) cannot be increased at the same rate as the amount of recorded data is going to increase.
A paradigm shift is needed for event simulation.
By 2030, when the High-Luminosity LHC~\cite{LHCSchedule} is expected to be in operation, about \SI{20}{\percent} of the computing resources are expected to be spent on event generation.
This emphasises the importance of projects such as \mgfgpu~\cite{mg4gpu}.

With the advent of SIMD-capable CPUs\footnote{Single Instruction Multiple Data} and GPUs, event generators can be improved by increasing data parallelism.
During the \mgfgpu project, \mg has been extended to support SIMD computations and GPUs~\cite{vCHEP2021,ICHEP2022,ACAT2022}.
Matrix-element computations as they are executed by \mg lend themselves particularly well to data parallelism, since the functions to evaluate matrix elements are identical for every event -- they are just run for different input data.
In addition, the matrix element computations are almost branch free, so vector computations using SIMD or GPU computations with low thread divergence can be employed with high efficiency.
\section{A GPU / SIMD backend for \mg}
\label{sec:GPU}

\begin{figure}[tb]
    \centering
    \subcaptionbox{Classic madevent workflow \label{fig:workflowClassic}}[0.25\textwidth]{
        \centering
        \includegraphics[height=0.3\textwidth, trim=0mm 2mm 0mm 1mm, clip=false]{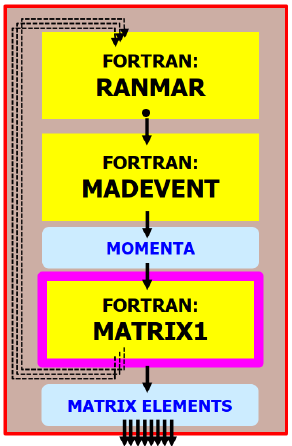}
    }%
    \hspace{0.049\textwidth}%
    \subcaptionbox{Standalone matrix-element workflow \label{fig:workflowStandalone}}[0.3\textwidth]{
        \centering
        \includegraphics[height=0.3\textwidth]{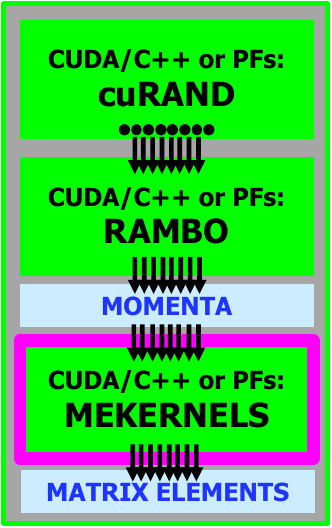}
    }%
    \hspace{0.049\textwidth}%
    \subcaptionbox{Multi-event madevent workflow \label{fig:workflowNew}}[0.25\textwidth]{
    \centering
        \includegraphics[height=0.3\textwidth]{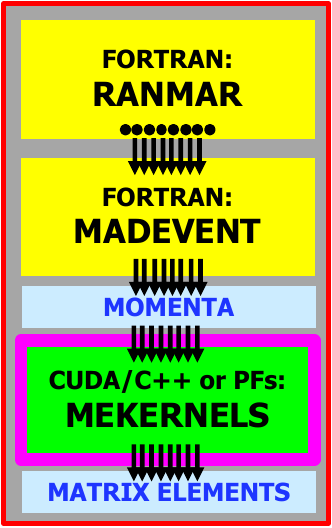}
    }
    \caption{Evolution of the madevent program. (\subref{fig:workflowClassic}) The classic madevent workflow ran random
    numbers, phase-space generation and matrix-element computation in a loop for one event at a time.
    (\subref{fig:workflowStandalone}) For \mgfgpu, a testbed with a simple, multi-event random number and phase-space generator was designed to test the parallel computation of matrix elements in \cpp, CUDA, or other portability frameworks.
    (\subref{fig:workflowNew}) Multi-event madevent, where parallel matrix-element computations from (\subref{fig:workflowStandalone}) can be used to benefit from SIMD vectorisation, multiple cores, or GPUs.
    }
    \label{fig:workflows}
\end{figure}

\mgfamc is a code generator to evaluate probability amplitudes, generate events, and compute cross sections.
For a given particle collision such as \eemumu\ or \process{pp}{t\bar{t}gg}, etc, the Fortran program ``madevent'' is generated and compiled, which contains a phase-space generator, a phase-space integrator, and a way to compute probability amplitudes (``matrix elements'') for a given collision, see \cref{fig:workflowClassic}.
The established \mg can generate the matrix-element code in Fortran, \cpp, and Python.
The \mgfgpu project is a plugin for \mg, where the \cpp matrix-element code generated by the original \mg was converted to \cpp and CUDA to evaluate probability amplitudes for batches of several thousands of events.
It relies on SIMD computations on CPUs and the even higher data parallelism on GPUs.

In addition to madevent, the plugin can generate a standalone program that only evaluates matrix elements without the phase-space integration step of madevent (\cref{fig:workflowStandalone}).
It relies on the simple phase-space generator ``RAMBO'', which is insufficient for LHC simulations, but ideal to quickly generate a batch of test events for throughput measurements.
This program was used to design, test and improve the initial \mgfgpu plugin~\cite{vCHEP2021}.
During a normal collision simulation, the madevent program calls the same code as the standalone program, so the latter can be used to measure the throughput of different approaches, or optimise the computation of matrix elements for specific hardware.
Using the standalone program, speedups of about \SI{250}{x} against the single-threaded Fortran amplitudes were achieved on the moderately complex process \process{gg}{t\bar{t}gg} using an NVidia Tesla V100 GPU in double precision mode, and \SIrange{4.6}{8.3}{x} using AVX2 and AVX512 vectorisation in double precision.
More details were shown on the ACAT 2022 conference~\cite{ACAT2022}.
For the complex process \process{gg}{t\bar{t}ggg}, the speedups are \SIlist{3.6;6.7;130}{x} for AVX2, AVX512, and CUDA on the V100 GPU.
This shows that a standalone computation of matrix elements can be sped up significantly using hardware accelerators, but for full event generation, these matrix elements have to be integrated into a larger framework.
This will be analysed in \cref{sec:hostSide}.
In single-precision mode, the matrix-element throughput can theoretically be doubled, but tracking the precision of the matrix elements showed that single-precision computations are not accurate enough to reliably integrate the phase space.

\subsection{Intel and AMD GPUs}
\begin{figure}[tb]
    \subcaptionbox{Scaling behaviour of matrix-element throughput for multiple parallel invocations of the standalone program \label{fig:SYCLScaling}}[0.4\textwidth]{
        \includegraphics[width=0.4\textwidth, trim=2mm 0mm 0mm 1mm, clip=false]{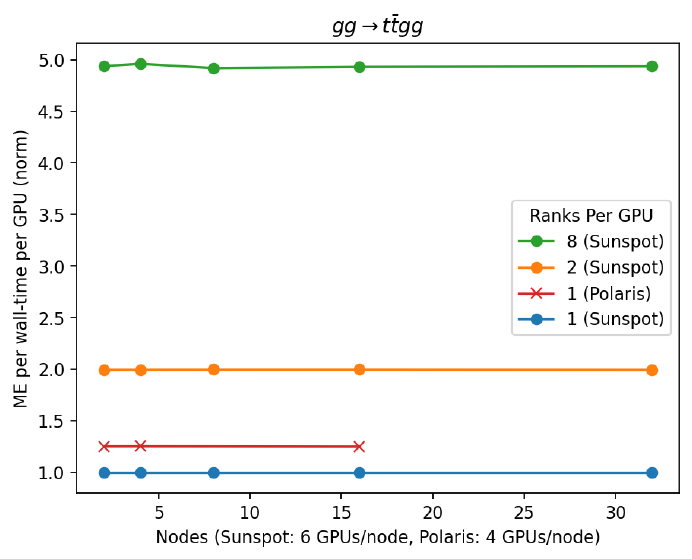}
    }%
    \hfill%
    \subcaptionbox{Throughput of the standalone program for different hardware \label{fig:SYCLComp}}[0.5\textwidth]{
        \includegraphics[width=0.5\textwidth, trim=14mm 2mm 0mm 0mm, clip=false]{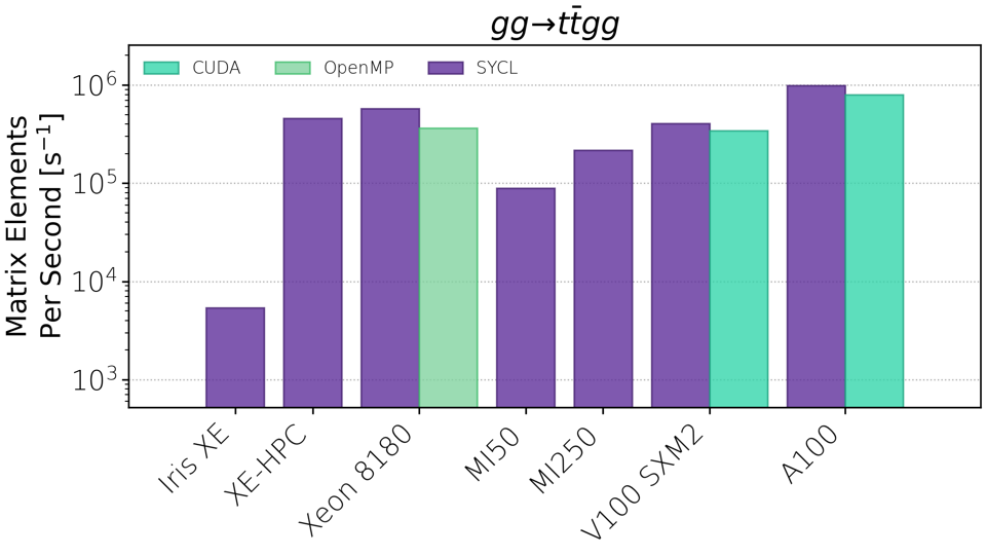}
    }%
    \caption{Throughput tests using the standalone program with a SYCL backend on the ``Sunspot'' testbed for the Aurora supercomputer at the Argonne National Laboratory. The program computes matrix elements for the process \process{gg}{t\bar{t}gg}. }%
    \label{fig:SYCL}%
\end{figure}

Based on the CUDA backend, other backends based on portability frameworks such as SYCL~\cite{SYCL}, Kokkos~\cite{Kokkos}, and Alpaka~\cite{Alpaka} were developed to test the feasibility of porting the matrix element code to different hardware.
CUDA code can be compiled for AMD GPUs with a small amount of changes, but Intel GPUs would be out of reach.
After initial progress with all backends, only the SYCL backend was continued due to limited resources.
\Cref{fig:SYCL} shows a scaling test and a throughput comparison on different hardware for the intermediate-complexity process \process{gg}{t\bar{t}gg}.
When starting multiple instances of the standalone program on multiple nodes of a high-performance computer, the throughput should scale with the number of processes.
This is confirmed in \cref{fig:SYCLScaling}.
\Cref{fig:SYCLComp} shows that matrix-element throughputs of \SIrange{1.E5}{1.E6}{\per \second} can be achieved with various GPUs from AMD, Intel and NVidia.
The classic Fortran matrix elements can be computed at a rate of \SI{3E3}{\per \second}, so a speedup by two orders of magnitude is achievable depending on the hardware used.
\section{Analysis of the full Madevent Workflow}
\label{sec:hostSide}

\begin{table}[htbp]
    \centering
    \sisetup{uncertainty-mode=separate,round-mode=uncertainty,round-precision=2}
    \begin{tabular}{l>{\footnotesize}l
        S[table-format=2.2 +- 1.2,table-align-text-post = true]
        S[table-format=2.2 +- 1.2,table-align-text-post = true]
        S[table-format=4.3 +- 1.3,table-align-text-post = true]}
    \toprule
    Process & \normalsize Matrix elm & {Total}  & {Momenta+} & {Matrix elm} \\
            &                             &          &  {unweight}  \\
    \midrule
    \eemumu
	&	Fortran	&	9.93	+-	0.05	s &	9.75	+-	0.05	s &	0.185	+-	0.001	s \\
\rowcolor{gray!10}	&	C++ AVX2	&	9.93	+-	0.02	s &	9.89	+-	0.02	s &	0.045	+-	0.001	s \\
\rowcolor{gray!10}	&		&	\color{ForestGreen} 1.00	+-	0.01	\x & \color{ForestGreen}	0.99	+-	0.01	\x & \color{ForestGreen}	4.12	+-	0.02	\x \\
	&	Cuda Tesla A100	&	10.33	+-	0.02	s &	10.32	+-	0.02	s &	0.008	+-	0.001	s \\
	&		&\color{ForestGreen}	0.96	+-	0.01	\x &\color{ForestGreen}	0.94	+-	0.01	\x &\color{ForestGreen}	24.3	+-	0.4	\x \\
\midrule													
\process{gg}{t\bar{t}gg}
    &	Fortran	&	106.6	+-	0.2	s &	4.55	+-	0.01   s	&	102.0	+-	0.2	s \\
\rowcolor{gray!10}
    &	C++ AVX2	&	29.01	+-	0.05	s &	4.56	+-	0.01 s	&	24.45	+-	0.04	s \\
\rowcolor{gray!10}
    &		&\color{ForestGreen}	3.67	+-	0.01	\x &\color{ForestGreen}	1.00	+-	0.01 \x	&\color{ForestGreen}	4.17	+-	0.01	\x \\
	&	Cuda Tesla A100	&	5.78	+-	0.01	s &	4.87	+-	0.01 s	&	0.91	+-	0.02	s \\
	&		&\color{ForestGreen}	18.44	+-	0.04	\x &\color{ForestGreen}	0.93	+-	0.01 \x	&\color{ForestGreen}	112.3	+-	2.1	\x \\
 \midrule
\process{gg}{t\bar{t}ggg}
    &	Fortran	&	2233.57	+-	1.85	s &	8.81	+-	0.07	s &	2224.764	+-	1.914	s \\
\rowcolor{gray!10}	&	C++ AVX2	&	697.17	+-	1.2	s &	8.71	+-	0.01	s &	688.4626	+-	1.1994	s \\
\rowcolor{gray!10}	&		& \color{ForestGreen}	3.20	+-	0.01	\x & \color{ForestGreen}	1.01	+-	0.01	\x & \color{ForestGreen}	3.23	+-	0.01	\x \\
	&	Cuda Tesla A100	&	27.78	+-	0.05	s &	9.12	+-	0.05	s &	18.66	+-	0.02	s \\
	&		& \color{ForestGreen}	80.402	+-	0.159	\x & \color{ForestGreen}	0.97	+-	0.01	\x & \color{ForestGreen}	119.226	+-	0.139	\x \\
    \bottomrule
    \end{tabular}

    \caption{Run times and speedup for different parts of the madevent executable for different processes.
    Madevent is invoked to produce $2^{18}$ weighted events, which are subsequently unweighted and written to disk.
    The Fortran parts of madevent such as phase-space sampling and unweighting run on the CPU, whereas matrix elements are either run on the CPU in Fortran, in \cpp with AVX2 extensions, or on the GPU in CUDA.
    The speedup is measured in comparison to the full-Fortran madevent program.
    CPU: AMD EPYC 7313, GPU: NVidia Tesla A100
    }
    \label{tab:madeventRunTimes}
\end{table}

In parallel with the development of the GPU plugin for \mg, madevent was converted to a multi-event interface.
Instead of generating random numbers, input and output particles, and the corresponding matrix element sequentially (event-by-event), random numbers and particles are now generated in batches of up to several thousand events.
This enables offloading matrix-element computations to the \cpp backend with SIMD acceleration, or to the CUDA backend for computation on a GPU as shown in \cref{fig:workflowClassic} and \cref{fig:workflowNew} in \cref{sec:GPU}.
The phase-space integration logic, as well as unweighting (that is selecting events that are written to the output file) remain in madevent. 
Inevitably, this leads to serial sections in madevent, so the total achievable speedup will according to Amdahl's law~\cite{Amdahl} be lower than the speedup achieved in the standalone program.

In the following, the integration of \cpp or CUDA matrix elements into madevent will be studied in more detail. 
The madevent executable is specific to each process being simulated, but it is mostly the matrix-element part that changes for different processes.
Before matrix elements on GPUs were implemented, madevent spent most CPU cycles in the matrix-element computation\footnote{except for very simple processes}, so this part was naturally optimised by the \mg developers.
The introduction of \cpp and CUDA matrix elements speeds up the matrix-element step by about a factor 4 for \cpp with AVX2 extensions and $> \SI{100}{\times}$ with CUDA, depending on the complexity of the process, so now the other parts of madevent also become relevant for the total run time, as shown in \cref{tab:madeventRunTimes}.
The table shows that for very simple processes with short matrix-element computations such as \eemumu, the achievable speedup is limited.
Due to an additional latency for transferring momenta and weights to a different computation backend, a speedup of the matrix elements doesn't speed up the full program, because the matrix elements run for an extremely short time.
For processes with intermediate complexity such as \process{gg}{t\bar{t}gg}, on the other hand, the speedup for matrix elements reaches \SI{4}{\x} for \cpp, and \SI{112}{\x} for CUDA.
The matrix elements are a significant part of the total run time here, so the overall speedup reaches almost \SI{4}{\x} with \cpp and \SI{18}{\x} with CUDA matrix elements.
For processes with high complexity, the matrix elements strongly dominate the run time, so \SI{3}{\x} and \SI{80}{\x} total speedup are achieved for \cpp and CUDA simulating the process \process{gg}{t\bar{t}ggg}.
Due to the Fortran parts of madevent taking an almost constant time independent of the matrix-element backend used, the total-program speedup increases with the complexity of the matrix elements being simulated.
This scaling is good for \mg’s users, since especially the complex processes with slow matrix elements will be of interest.

The sequential parts of madevent that limit the speedup will be analysed in the following.
\Cref{fig:flame} shows where madevent for \process{gg}{t\bar{t}gg} spends CPU cycles, depending on what type of matrix elements are used.
The Fortran version spends most of the CPU cycles in the function \texttt{matrix1\_}, \cref{fig:flameFortran}, which computes the matrix elements.
In \cref{fig:flameCUDA} on the other hand, the matrix elements are barely visible due to the CUDA acceleration.
The matrix elements are computed in the deep call stack on the right of the figure, which is so narrow that function names could not be shown.
With CUDA acceleration, the bulk of the run time is spent in unweighting (\texttt{unwgt\_}), event I/O (\texttt{sample\_put\_point\_}), phase-space sampling (\texttt{sample\_full\_}), and the evaluation of the Parton Distribution Functions (PDFs, \texttt{lh\_polint\_}).

Analysing these parts further, it became evident that the madevent unweighting algorithm limits the total speedup, since it requires writing events to temporary storage if they cannot be discarded immediately.
To sample unweighted events from a batch of weighted events, the maximum event weight of each batch needs to be known.
However, since madevent was designed to iterate through an event sample one by one, it used to employ a running maximum.
It therefore temporarily accepted events, and wrote them to temporary storage, but most of these were discarded later once the global maximum was known.
These events were written to a file, and had to be read again once the decision could be made to accept or discard them (cf.\ \texttt{write\_event\_} and \texttt{read\_event\_} in \cref{fig:flameCUDA}).

Given that madevent has been converted to a multi-event interface to enable GPU and vectorised computations, c.f.\ \cref{fig:workflowNew}, the maximum weight of a set of events can be computed in one go instead of using a running maximum.
To do this, weights from Jacobian terms and PDFs are transferred to the GPU, and multiplied with the matrix elements that are already in GPU memory to compute the full event weight.
The maximum of these weights is computed in parallel, and transferred back to the host.
The knowledge of this weight improves the efficiency of the unweighting steps, as low-weight events can be discarded at a higher rate, and fewer events have to be written into or retrieved from temporary storage.
For \process{gg}{t\bar{t}gg}, for example, the default unweighting of madevent temporarily stored \num{89923} out of \num{278506} events, but the final sample consisted of \num{870} events.
Using the GPU-assisted unweighting, \num{1475} out of \num{278506} were retained, and the final unweighted sample consisted of \num{1053} unweighted events.
The amount of events that were stored temporarily but ultimately discarded was therefore reduced to \SI{5}{\permille}, and the number of events produced by madevent was increased.
In \cref{tab:gpuAssistedUnweighting}, the speedup by employing GPU matrix elements with batch unweighting is shown.
In contrast to \cref{tab:madeventRunTimes}, the zero-width t-channel mode is off, so absolute run times cannot be compared.
By employing GPU-assisted unweighting, the slowdown that is normally incurred by transferring momenta from madevent to an accelerated matrix-element backend is compensated for, and a speedup of the momenta+unweight step is achieved.
\begin{table}[htb]
    \centering
    \sisetup{uncertainty-mode=separate,round-mode=uncertainty,round-precision=2}
    \begin{tabular}{l>{\footnotesize}l
        S[table-format=4.2 \pm 1.2,table-align-text-post = true]
        S[table-format=2.2 \pm 1.2,table-align-text-post = true]
        S[table-format=3.2 \pm 1.2,table-align-text-post = true]}
    \toprule
    Process & \normalsize Matrix elm & {Total}  & {Momenta+unweight} & {Matrix elm} \\
    \midrule
    \process{gg}{t\bar{t}gg}
    	&	Fortran	&	108.1	+-	0.27	s &	6.27	+-	0.41	s &	101.8357	+-	0.1361	s \\
\rowcolor{gray!10}	&	C++ AVX2	&	31.08	+-	0.01	s &	6.88	+-	0.01	s &	24.20	+-	0.02	s \\
\rowcolor{gray!10}	&		& \color{ForestGreen}	3.48	+-	0.01	\x & \color{ForestGreen}	0.91	+-	0.06	\x & \color{ForestGreen}	4.21	+-	0.01	\x \\
	&	Cuda Tesla A100	&	5.32	+-	0.03	s &	4.67	+-	0.02	s &	0.66	+-	0.02	s \\
	&		& \color{ForestGreen}	20.320	+-	0.125	\x & \color{ForestGreen}	1.34	+-	0.09	\x & \color{ForestGreen}	155.379	+-	2.67	\x \\
     \bottomrule
    \end{tabular}
    \caption{Run times and speedup for the madevent executable with batch unweighting.
    In contrast to \cref{tab:madeventRunTimes}, the zero-width t-channel mode is deactivated, so absolute run times cannot be compared between the two tables.
    When GPU matrix elements are used, the unweighting of each batch of \num{16384} events completes faster and with higher efficiency.
    }
    \label{tab:gpuAssistedUnweighting}
\end{table}

\begin{figure}[p]
    \centering
    \subcaptionbox{Fortran-only execution \label{fig:flameFortran}}[\textwidth]{
        \adjustbox{trim=0cm 0cm 0cm 3cm, clip=true}{%
            \includesvg[width=\textwidth]{figures/flame_ggttgg.mad_FOR_noMatrix1Counter}
        }
    }
    \subcaptionbox{Fortran + CUDA execution \label{fig:flameCUDA}}[\textwidth]{
        \adjustbox{trim=0cm 0cm 0cm 2.5cm, clip=true}{%
            \includesvg[width=\textwidth]{figures/flame_GPU_oldUnweighting}
        }
    }
    \subcaptionbox{Fortran + CUDA execution with GPU-assisted unweighting \label{fig:flameCUDAUnweight}}[\textwidth]{
        \adjustbox{trim=0cm 0cm 0cm 3cm, clip=true}{%
           \includesvg[width=\textwidth]{figures/flame_maxWeightKernel}
        }
    }
    \caption{
        Flamegraph analysis~\cite{flamegraph} of CPU cycles spent in three madevent runs for the process \process{gg}{t\bar{t}gg}.
        The fortran-only version \textbf{(\subref{fig:flameFortran})} spends most of the time in the matrix-element computation (\texttt{matrix1\_}).
        In \textbf{(\subref{fig:flameCUDA})}, this function is sped up about \SI{170}{x} using a GPU.
        Now, phase-space sampling, unweighting, and PDF evaluation dominate the run time.
        In \textbf{(\subref{fig:flameCUDAUnweight})}, the unweighting step was improved in addition to using GPU matrix elements by employing more efficient, GPU-assisted unweighting.
        This results in a speedup of \SI{1.3}{x} on the host side.
        The total run time is now dominated by phase-space sampling (\texttt{sample\_full\_}) and PDF evaluation (\texttt{lh\_polint\_}).
    }
    \label{fig:flame}
\end{figure}

The impact on madevent is shown in \cref{fig:flameCUDAUnweight}.
A large reduction of \texttt{write\_event\_} and \texttt{read\_event\_} calls is observed with respect to \cref{fig:flameCUDA}, and the madevent run time without the matrix-element part is reduced from \SIrange{6.3}{4.7}{s} as shown in \cref{tab:gpuAssistedUnweighting}.
The number of events in temporary storage, unweighting efficiency, and the achievable speedup with GPU-assisted unweighting depend on the process being simulated, but in all cases, the direct computation of the maximum reduces the number of events to be stored.

Given that madevent can be used in a multi-event workflow also without GPUs, a similar unweighting strategy could be employed if Fortran and \cpp matrix elements are used.
This algorithmic improvement will be tested in madevent.
It should be stressed that the main reason for a speedup is not the usage of the GPU to compute the maximum weight, but it is the knowledge of the maximum event weight for a larger batch of events.
This leads to better accept/reject decisions, and fewer events are written or read to/from files.

With GPU matrix elements and GPU-assisted batch unweighting, the run times to compute \process{gg}{t\bar{t}gg} are now dominated by the PDF evaluation (\texttt{lh\_polint\_}) and the phase-space sampling (\texttt{sample\_full\_}) steps as shown in \cref{fig:flameCUDAUnweight}.
These could be reduced further, by e.g.\ employing accelerated PDF libraries or by reworking the phase-space algorithm of madevent.
Currently, though, the focus of the \mgfgpu project is on releasing the plugin for testing by the LHC experiments.
\section{Summary and Outlook: Releasing \mgfgpu to \mgfamc's users}

The \mgfgpu plugin was shown to significantly speed up matrix-element computations, especially for processes with many final-state particles.
These are computationally expensive to simulate, so this plugin should be a default choice for \mg if high statistics are required.
Even if \mg users don’t have access to GPU resources, they can use the plugin to make use of SIMD computations, which are supported by all modern CPUs.

At the time of the conference, however, \mg users were not able to generate their own madevent executables accelerated by the \mgfgpu plugin.
First ``gridpacks'' had been produced by the \mgfgpu developers with CUDA and vectorised \cpp backends, but this required a non-standard \mg.
For a gridpack, \mg pre-samples the phase space of a given process, and optimises the integration grid.
This ensures that the sampling frequency of phase-space regions matches their probability density.
This pre-sampled grid is written into files, and packed into archives together with the madevent executables for each subprocess.
These archives can be used to run larger batch jobs that generate the desired number of collision events.
For a few select processes, the ATLAS and CMS experiments were able to confirm that the gridpacks can benefit from GPU acceleration.
The gridpacks were only tested -- and equality of results against \mg verified -- for a limited number of subprocesses such as \eemumu\  and \process{gg}{t\bar{t}\,0\mathit{\textnormal{-}}3g}.

The \mgfgpu plugin currently only supports leading-order QED and QCD processes, because running couplings in weak interactions and SUSY or BSM processes will require a more elaborate treatment in the GPU backend.
These are on the plan of work.
Processes with multiple subprocesses, such as proton-proton collisions \process{pp}{t\bar{t}\,0\mathit{\textnormal{-}}2j}, that is none to two additional jets, are being tested currently.

Lastly, work is underway to improve the integration into \mgfamc.
In the future, \mg users will be able to check out the \mgfgpu plugin, and run the generation of the matrix-element code using the native \mg interface.
We hope that this work reduces waiting times of \mg users, and helps to alleviate the pressure on scarce computing resources in light of growing datasets during Run 3 and the High-Luminosity LHC.

\bibliography{main.bib}

\end{document}